# Nonrelativistic quantum theory of non-dipole effects in a multielectron atom: radiation transition operator

Alexey N. Hopersky, Alexey M. Nadolinsky, and Rustam V. Koneev

*Rostov State Transport University, 344038, Rostov-on-Don, Russia*
*E-mail*: qedhop@mail.ru, amnrnd@mail.ru, koneev@gmail.com

---

**Abstract.** In the work of Hopersky A.N. *et al.*, arXiv: 2609.33494v1 [physics. atom-ph] in the Loudon integral representation, the general analytical structure of the complete operator of a multipole radiation transition taking into account non-dipole effects for a multielectron atom is established. In this **Letter** this structure is converted to the sum only for all multipoles.

---

Without repeating the details of the construction and the corresponding description of the designations, we will present the general result of the work [1] (see formula (10)) for the complete operator of a multipole radiation transition in a multielectron atom:

$$H_e = i\frac{e}{c}\sum_{\vec{k}\rho}\sum_{p}\sum_{l=0}^{\infty}\sum_{LM}\omega_k(-1)^{p+M}C_{-p}^{(1)}(\vec{e}_{\vec{k}\rho})\cdot i^l\cdot(2l+1)(2L+1)\cdot$$
$$\cdot[(-1)^l a^+_{\vec{k}\rho}-a^-_{\vec{k}\rho}]\cdot\begin{pmatrix}1 & l & L\\ 0 & 0 & 0\end{pmatrix}\begin{pmatrix}1 & l & L\\ p & 0 & -M\end{pmatrix}\cdot\int_0^1 d\lambda\cdot\sum_{n=1}^{N} r_n j_l(\lambda k r_n)C_M^{(L)}(\vec{e}_n). \quad \textbf{(1)}$$

In operator (1) the term with $L=0$ ($M=0$) is *absent* in the scheme of the proposed experiment on the interaction of X-rays with an atom: $\vec{k}\parallel OZ$, $\vec{e}_{\vec{k}\rho}\perp OZ$, $(\vec{k}\cdot\vec{e}_{\vec{k}\rho})=0$.

In operator (1), we perform the summation on quantum numbers $l=L\pm1$ (see conditions on the top line of the $3j$–Wigner symbol) and $p=\pm1$ ($p=0$ *disappears* due to equality for the spherical function $C_0^{(1)}(\vec{e}_{\vec{k}\rho})=0$ at $\vec{k}\parallel OZ$). When summing over the $M$–quantum number, it is taken into account that $M=p$ by virtue of the appearance of the $\delta_{pM}$ – Kronecker symbol (see the condition on the bottom line of the $3j$–Wigner symbol: $p-M=0$). Thus, from the full set $M=-L,\dots,L$ only the values $M=\pm1$ are "cut" for any values of multipole $L\geq1$.

**1. Summation by $l$.** Consider the result for $3j$ –Wigner symbol with a bottom string of zeros [2]:

$$\begin{pmatrix}j_1 & j_2 & j\\ 0 & 0 & 0\end{pmatrix}=(-1)^g\sqrt{\frac{(2g-2j_1)!(2g-2j_2)!(2g-2j)!}{(2g+1)!}}\frac{g!}{(g-j_1)!(g-j_2)!(g-j)!}, \quad (2)$$

where $j_1+j_2+j=2g$, $g$ – integer. Then we have:

$$\begin{pmatrix}1 & L+1 & L\\ 0 & 0 & 0\end{pmatrix}=(-1)^{L+1}\sqrt{\frac{L+1}{(2L+1)(2L+3)}}, \quad (3)$$

$$\begin{pmatrix}1 & L-1 & L\\ 0 & 0 & 0\end{pmatrix}=(-1)^{L}\sqrt{\frac{L}{(2L-1)(2L+1)}}. \quad (4)$$

Let's enter the notations:

$$\alpha_L=\sqrt{(L+1)(2L+3)},\quad \beta_L=\sqrt{L(2L-1)}, \quad (5)$$

$$B_{L\pm1}^{(p)}=\sum_{n=1}^{N}C_p^{(L)}(\vec{e}_n)\int_0^1 d\lambda\cdot r_n\cdot j_{L\pm1}(\lambda k r_n), \quad (6)$$

$$\gamma_{\vec{k}\rho}^{(L)} = \left\lfloor (-1)^L a_{\vec{k}\rho}^{+} + a_{\vec{k}\rho}^{-} \right\rfloor. \tag{7}$$

Then operator (1) takes the form:

$$\begin{gathered} H_e = \frac{e}{c} \sum_{\vec{k}\rho} \sum_{p} \sum_{L=1}^{\infty} (-1)^{L+1} C_{-p}^{(1)}(\vec{e}_{\vec{k}\rho}) \sqrt{2L+1} \cdot \gamma_{\vec{k}\rho}^{(L)} \cdot i^L \cdot \\ \cdot \left\{ \alpha_L \cdot B_{L+1}^{(p)} \begin{pmatrix} 1 & L+1 & L \\ p & 0 & -p \end{pmatrix} + \beta_L \cdot B_{L-1}^{(p)} \begin{pmatrix} 1 & L-1 & L \\ p & 0 & -p \end{pmatrix} \right\}. \end{gathered} \tag{8}$$

**2. Summation by** $p$**.** Consider the result for the $3j$–Wigner symbol [2]:

$$\begin{pmatrix} j_1 & j_2 & j \\ m_1 & m_2 & m \end{pmatrix} = (-1)^{j_1+j_2+j} \cdot \begin{pmatrix} j_1 & j_2 & j \\ -m_1 & -m_2 & -m \end{pmatrix}. \tag{9}$$

Then we have:

$$\begin{pmatrix} 1 & L\pm 1 & L \\ -1 & 0 & 1 \end{pmatrix} = \begin{pmatrix} 1 & L\pm 1 & L \\ 1 & 0 & -1 \end{pmatrix}, \tag{10}$$

Let us also take into account that in case of $\vec{k} \parallel OZ$ spherical function $C_{\pm 1}^{(1)}(\vec{e}_{\vec{k}\rho}) = \mp\sqrt{3/8\pi}$ and enter the notations:

$$A_{L\pm 1} = B_{L\pm 1}^{(1)} - B_{L\pm 1}^{(-1)}. \tag{11}$$

Then operator (8) takes the form:

$$\begin{gathered} H_e = -\frac{e}{c} \sqrt{3/8\pi} \sum_{\vec{k}\rho} \sum_{L=1}^{\infty} (-1)^L \sqrt{2L+1} \cdot \gamma_{\vec{k}\rho}^{(L)} \cdot \\ \cdot \left\{ \alpha_L \cdot A_{L+1} \cdot \begin{pmatrix} 1 & L+1 & L \\ 1 & 0 & -1 \end{pmatrix} + \beta_L \cdot A_{L-1} \cdot \begin{pmatrix} 1 & L-1 & L \\ 1 & 0 & -1 \end{pmatrix} \right\}. \end{gathered} \tag{12}$$

To calculate the $3j$ –Wigner symbols in (12), we take into account the result [2]:

$$\begin{pmatrix} j_1 & j_2 & j \\ m_1 & m_2 & -m \end{pmatrix} = (-1)^{j_1-j_2+m} \cdot \sqrt{\frac{(2j_1)!(2j_2)!(j+m)!(j-m)!}{(2j+1)!(j_1+m_1)!(j_1-m_1)!(j_2+m_2)!(j_2-m_2)!}}, \tag{13}$$

where $j = j_1 + j_2$ and $m = m_1 + m_2$. Then we have:

$$\begin{pmatrix} 1 & L+1 & L \\ 1 & 0 & -1 \end{pmatrix} = (-1)^{L+1} \sqrt{\frac{L}{2(2L+1)(2L+3)}}, \tag{14}$$

$$\begin{pmatrix} 1 & L-1 & L \\ 1 & 0 & -1 \end{pmatrix} = (-1)^{L+1} \sqrt{\frac{L+1}{2(2L-1)(2L+1)}}. \tag{15}$$

In (14) the permutation of the second and third columns of the $3j$– Wigner symbol is taken into account. Back to the operator (12), taking into account (5), (11), (14) and (15) we finally get:

$$H_e = \sum_{\vec{k}\rho} \sum_{L=1}^{\infty} i^L \sqrt{\frac{L(L+1)}{2}} \cdot \gamma_{\vec{k}\rho}^{(L)} \cdot Q^{(L)}, \tag{16}$$

$$Q^{(L)} = \frac{e}{c} \sqrt{3/8\pi} \cdot \sum_{n=1}^{N} (C_1^{(L)}(\vec{e}_n) - C_{-1}^{(L)}(\vec{e}_n)) \cdot \Phi_n^{(L)}, \tag{17}$$

$$\Phi_n^{(L)} = \frac{2L+1}{k} \cdot \int_0^{r_n} \frac{dx_n}{x_n} \cdot j_L(kx_n). \tag{18}$$

In the operator (16) $k = 2\pi/\lambda_\omega$, $\lambda_\omega$ – photon wavelength, the recurrence relation for spherical Bessel functions is taken into account [3]

$$j_{L+1}(z) + j_{L-1}(z) = \frac{2L+1}{z} j_L(z) \tag{19}$$

and indicated $x_n = \lambda r_n$.

So, operator (1) is converted to sum only for all multipoles:

$$H_e = \sum_{L=1}^{\infty} H_e^{(L)} . \quad (20)$$

In doing so, the value $L$ determines the appropriate selection rules for quantum numbers in probability amplitudes of a multipole transition. Of course, the analytic structure of the $\Phi_n^{(L)}$ – operator of (18) is important, first of all, for constructing the probability amplitudes of a multipole transition between single-electron *continuous*-spectrum states. In fact, in this case, the dipole approximation *loses its physical meaning* due to the non-localizability of these states. From (16), in particular, for the *dipole* transition ($L = 1$) we have:

$$H_e^{(1)} \equiv H_e^{(D)} = i \sum_{\vec{k}\rho} \gamma_{\vec{k}\rho}^{(1)} \cdot Q^{(1)} , \quad (21)$$

$$Q^{(1)} = \frac{e}{c} \sqrt{3/8\pi} \cdot \sum_{n=1}^{N} (C_{-1}^{(1)}(\vec{e}_n) - C_1^{(1)}(\vec{e}_n)) \cdot \Phi_n^{(1)} , \quad (22)$$

$$\gamma_{\vec{k}\rho}^{(1)} = \omega_k (a_{\vec{k}\rho}^{+} - a_{\vec{k}\rho}^{-}) , \quad (23)$$

$$\Phi_n^{(1)} = \frac{3}{k} \cdot \int_0^{r_n} \frac{dx_n}{x_n} \cdot j_1(kx_n) . \quad (24)$$

The result for the operator (21) reproduces the result (16) of the work [1].

The analytical structure (16) *summarizes* the formulas (7), (8) and (9) of the work [4] obtained in the approximation of a "split" electron of a *continuous* spectrum. Thus, in Loudon's integral representation [5] (see equation 5.35 for the electrical part of the total interaction operator), result (16) defines the multipole structure of the total radiation transition operator in the non-relativistic quantum theory of non-dipole effects in a multielectron atom. Further construction of the probability amplitudes of multipole transitions within the framework of the presented theory is carried out by the methods of the theory of irreducible tensor operators, the theory of non-orthogonal orbitals and the algebra of production (annihilation) operators. As an example of the implementation of result (24) in a simple approximation of substitution [1]

$$\Phi_n^{(1)} \to D_n = \begin{cases} r_n, & r_n \in [0; q), \\ q, & r_n \in [q; \infty), \end{cases} \quad (25)$$

where $q = (3/8)\lambda_\omega$, note the discovery of a *giant* non-dipole effect in the theoretical description of the generalized cross-sections of the two-photon double ionization of the *K*-shell of neon-like ($Fe^{16+}$) [6] and beryllium-like ($Ti^{18+}$, $Fe^{22+}$, $Zn^{26+}$) [7] atomic ions.

---